\documentclass[a4paper,11pt,twoside]{article}

\usepackage{amsmath}
\usepackage{amssymb}
\usepackage{epsfig}

\newcommand{\dm}[1]{{\Delta m^2_{#1}}}

\newcommand{\Fig}[1]{Figure~\ref{fig:#1}}

\newcommand{\Eq}[1]{Eq.~(\ref{eq:#1})}
\newcommand{\eq}[1]{eq.~(\ref{eq:#1})}

\newcommand{\interskip}{\bigskip}
\newcommand{\tab}{\theta_{12}}
\newcommand{\tbc}{\theta_{23}}
\newcommand{\tac}{\theta_{13}}
\newcommand{\ord}[1]{\mathcal{O}\left( #1 \right)}

\newcommand{\capdef}{}
\newcommand{\mycaption}[2][\capdef]{\renewcommand{\capdef}{#2}%
        \caption[#1]{{#2}}} 
\makeatletter
\renewcommand{\fnum@table}{\textbf{\tablename~\thetable}}
\renewcommand{\fnum@figure}{\textbf{\figurename~\thefigure}}
\makeatother
\makeatletter
\renewcommand{\section}{\@startsection{section}{1}{0em}%
        {-3.25ex \@plus -1ex \@minus -.2ex}%
        {1.5ex \@plus.2ex}%
        {\normalfont\bfseries}}
\renewcommand{\subsection}{\@startsection{subsection}{2}{0em}%
        {-3.25ex\@plus -1ex \@minus -.2ex}%
        {1.5ex \@plus .2ex}%
        {\normalfont\bfseries}}
\renewcommand{\subsubsection}%
        {\@startsection{subsubsection}{3}{0em}%
        {-3.25ex\@plus -1ex \@minus -.2ex}%
        {1.5ex \@plus .2ex}%
        {\normalfont\itshape}}
\makeatother

\newcommand{\teab}{\theta^e_{12}}
\newcommand{\teabc}{\theta^{e_c}_{12}}
\newcommand{\htac}{\hat\theta_{13}}

\newcommand{\CERN}{CERN, Department of Physics, Theory Division \\
  CH--1211 Geneva 23, Switzerland}

\newcommand{\preprintdate}{February 2004}
\newcommand{\preprintnumber}{CERN-TH/2004-039}
 
\newcommand{\titletext}{Charged lepton contributions to the solar \\
  neutrino mixing and $\theta_{13}$} 
\newcommand{\authortext}{\large Andrea Romanino
\medskip\\\em\normalsize \CERN}
\newcommand{\abstracttext}{A charged lepton contribution to the solar
  neutrino mixing induces a contribution to $\tac$, barring
  cancellations/correlations, which is independent of the model
  building options in the neutrino sector. We illustrate two robust
  arguments for that contribution to be within the expected
  sensitivity of high intensity neutrino beam experiments. We find
  that the case in which the neutrino sector gives rise to a maximal
  solar angle (the natural situation if the hierarchy is inverse)
  leads to a $\tac$ close to or exceeding the experimental bound
  depending on the precise values of $\tab$, $\tbc$, an unknown phase
  and possible additional contributions. We finally discuss the
  possibility that the solar angle originates predominantly in the
  charged lepton sector. We find that the construction of a model of
  this sort is more complicated. We comment on a recent example of
  natural model of this type.}

\title{
\normalsize
\begin{tabular}[t]{l}
\preprintdate\end{tabular}
\hspace*{\fill}
\begin{tabular}[t]{l}\preprintnumber\end{tabular}
\vspace{3\baselineskip}\\\Large\bfseries\titletext\bigskip}
\author{\begin{minipage}[t]{0.8\textwidth}
\normalsize\centering\authortext
\end{minipage}}
\date{}

\begin{document}

\bigskip
\maketitle
\begin{abstract}\normalsize\noindent
\abstracttext
\end{abstract}\normalsize\vspace{\baselineskip}



\section{Introduction}

The steady, remarkable progress we have witnessed in recent years in
experimental neutrino physics has enabled a significant advance in our
understanding of the lepton sector. The case for three neutrino
oscillations is now compelling (although a full oscillation pattern
still has to be observed) and the peculiar neutrino mass and mixing
pattern observed represents a non-trivial handle on their origin. It
is then natural to wonder what the understanding we gained implies for
the value of the observables still to be measured, in particular
$\tac$. A major part of the rich neutrino experimental program
available and partially under way will in fact focus on measuring that
mixing angle, which has also important implications for leptonic CP
violation and astrophysics. While many models do provide predictions
for $\tac$, the number of possibilities is high enough to make almost
any value of $\tac$, from zero to the present bound, compatible with
some model. In this letter, we will therefore try to overcome the bulk
of the model dependence in those predictions by focusing on general
mechanisms leading to calculable contributions to $\tac$.

While both the neutral and charged lepton sectors contribute to lepton
mixing, most of the uncertainties in model building come from the
neutrino sector. This is because the light neutrino mass matrix is
still less constrained than the charged lepton one and, more
important, because the origin of its smallness introduces additional
degrees of freedom. For example, in the case of the type I see-saw
mechanism, the light neutrino matrix is determined by two independent
mass matrices\footnote{One way of getting oriented in the jungle of
  model building possibilities in the neutrino sector is by
  considering minimal models~\cite{Barbieri:03a}. See also the general
  arguments in~\cite{Akhmedov:99a,Barr:00a,Feruglio:02a,Lavignac:02a}.
  In~\cite{FPR} the assumption is made that the neutrino contribution
  to the mixing matrix is bimaximal. In some cases the model building
  in the neutrino and charged lepton sector can be closely
  related~\cite{Kuchimanchi:02a}.}. We will therefore concentrate on
contributions to $\tac$ that are independent of the model building in
the neutrino sector and rely instead on properties of the charged
lepton sector\footnote{This contribution arises in a number of
  explicit models, in some cases providing a precise prediction for
  $\tac$, see
  e.g.~\cite{Barbieri:99a,Barr:00a,Ibarra:03a,Lebed:03a,FPR}.  We
  focus here on the general features of the effect.}. In particular,
we identify a contribution to $\tac$ induced by the charged lepton
contribution to the solar mixing angle, which arises in the absence of
correlations in the charged lepton mass matrix. We discuss two
motivated expectations for the size of the latter contribution, both
leading to values of $\tac$ likely to be within the reach of high
intensity conventional neutrino beam experiments. In particular, we
consider the case in which the neutrino sector gives rise to a maximal
solar angle.  This is typically the case in models with inverse
hierarchical neutrinos~\cite{King:00a}. In this case, the charged
lepton sector must account for the observed deviation of $\tab$ from
$45^\circ$, which in turn leads to a contribution to $\tac$ close to
the present experimental bound~\cite{King:00a,King:02a}. We provide a
simple analytical expression for the latter involving one physical
phase and we plot the distribution of the corresponding numerical
expectation taking into account the uncertainty on the mixing
parameters and possible additional contributions~\cite{Me}.  Indeed,
since additional contributions to $\tac$ may be present, accidentally
cancelling the piece controlled by the charged lepton sector, the
expectations we find should be considered as lower limits on $\tac$,
uncertain by a factor of order one. A condition for a non accidental
cancellation is discussed in~\cite{AFM}.

We also discuss the possibility that the charged lepton contribution
to the solar mixing angle is large and accounts for most of it. We
first consider the case in which the entries in the charged lepton
mass matrix are uncorrelated and show that i) the induced contribution
to $\tac$ is well beyond the experimental bound (the ``$\tac$ tuning
problem'') and that ii) in SU(5) grand unified models, the electron
mass (or the up quark mass) gets a contribution way larger than the
observed value (the ``$m_e$ tuning problem'', numerically more
relevant). This makes the construction of a natural model of this sort
considerably more complicated~\cite{Me}. We then discuss the case in
which correlations are present in the charged lepton mass matrix. In
particular, we consider the possibility, recently pointed out
in~\cite{AFM}, that such correlations arise from a non-accidental
physical mechanism. A condition on the charged lepton mass matrix
sufficient to naturally solve the $\tac$ tuning problem has been
identified in~\cite{AFM}. We show that an additional condition is
necessary to ensure that the $m_e$ tuning problem is also solved in
SU(5) models. The mechanism illustrated in~\cite{AFM} may easily
account for the latter condition as well.

\section{Charged lepton rotations and $\tac$}

The neutrino mixing matrix $U$ is the combination of two unitary
matrices, $U_e$ and $U_\nu$, entering the diagonalization of the
charged lepton and neutrino mass matrices $m_E$ and
$m_\nu$\footnote{We use a convention in which the left-handed fields
  lie on the right hand side of $m_E$.},
\begin{equation}
  \label{eq:U}
  U = U_e U^\dagger_\nu \text{, where} \quad m_E = U^T_{e_c}
  m^{\text{diag}}_{E} U_e \text{,} \quad m_\nu = U_\nu^T
  m^{\text{diag}}_\nu U_\nu \,.
\end{equation}
The decomposition of $U$ into $U_e$ and $U_\nu$ is of course not
physical in the (minimally extended) Standard Model (SM)\footnote{An
  $SU(2)_L$-invariant transformation in flavour space can equally well
  rotate the physical mixing in the neutrino and in the charged lepton
  sector.}.  On the other hand, any attempt at investigating the
origin of the fermion flavour structure involves physics beyond the SM
that identifies a privileged basis in flavour space. This is the basis
in which the entries of the fermion mass matrices are most simply
related to the physics originating them and in which possible
correlations among the entries in $m_E$, $m_\nu$ have to be considered
accidental or related to symmetries or other physical mechanisms of
the underlying theory.

We can then wonder if lepton mixing, and in particular the two large
mixing angles observed in atmospheric, solar and terrestrial neutrino
experiments, originate from the neutrino or the charged lepton sector.
It is well known that the atmospheric angle $\tbc$ can equally well
originate from the neutrino~\cite{Smirnov:93b} or the charged
lepton~\cite{Albright:98a} sector. The solar angle can in principle
also come from both sectors.  However, as we will see later on,
cancellations or correlations in the charged lepton mass matrix are
required in the case in which the origin of the solar angle is in the
lepton sector. For the time being, we consider the case in which the
entries of $m_E$ are not correlated and show that in general a
left-handed 12-rotation in the charged lepton sector $\teab$ induces a
contribution to $\tac$. This derivation is closely related to that
in~\cite{Barr:00a}.

Barring correlations in the charged lepton mass matrix
$m_E$\footnote{Democratic models are an example in which such
  correlations are not accidental.}, the hierarchy of charged lepton
masses translates in a hierarchical structure of $m_E$:
\begin{equation}
  \label{eq:hierarchy}
  \begin{vmatrix}
    m_{11} & m_{12} & m_{13} \\
    m_{21} & m_{22} & m_{23} \\
    m_{31} & m_{32} & m_{33}
  \end{vmatrix}^E \ll
  \begin{vmatrix}
    m_{22} & m_{23} \\
    m_{32} & m_{33}
  \end{vmatrix}^E m_\tau \ll |m_{33}|^E m^2_\tau \;.
\end{equation}
Note that the condition~(\ref{eq:hierarchy}) is compatible with
asymmetrical matrices and in particular with the atmospheric angle
originating from $m_E$. 

The charged lepton mass matrix can be approximately diagonalized by
the subsequent diagonalization of his $2\times 2$ blocks. As a
consequence of~(\ref{eq:hierarchy}), the diagonalization of the
heaviest 23 block should be performed first, so that
\begin{equation}
  \label{eq:order}
  U_e = U^e_{12} U^e_{13} U^e_{23} \;,
\end{equation}
where $U^e_{ij}$ is a complex rotation in the $ij$ block. The
ordering~(\ref{eq:order}) ensures that the mixing parameters are
directly related to the entries of $m_E$ and are therefore
independent, barring correlations already present in $m_E$ or induced
by the first steps of diagonalization. For our purposes, it is
sufficient to consider the two cases $U^e_{12}\neq \mathbf{1}$,
$U^e_{13} = \mathbf{1}$ and $U^e_{13}\neq \mathbf{1}$, $U^e_{12} =
\mathbf{1}$. Since the two possibilities are actually equivalent (up
to a relabelling of the first two rows of $m_E$) in the $\tbc=\pi/4$
limit, in this paper we will consider the $U^e_{12}\neq \mathbf{1}$,
$U^e_{13} = \mathbf{1}$ case only.

When combining $U_e$ and $U_\nu$ in the physical neutrino mixing
matrix, the $U^e_{12}$ rotation ends up on the left-hand side of $U$:
\begin{equation}
  \label{eq:Lside} 
  U = U^e_{12} \hat U ,
\end{equation}
where $\hat U = U^e_{23} U^\dagger_\nu$ and the standard
parameterization can be used for $U_\nu$. Note that, wherever it comes
from, the atmospheric angle resides in $\hat U$. Now, in order to read
the value of the solar angle $\tab$ from \eq{Lside} we have to write
$U$ in the parameterization that defines $\tab$, in which the 12
rotation lies on the right-hand side: $U = U_{23}U_{13}U_{12}$. This
means that $U^e_{12}$ has to be commuted with $\hat U$ and in
particular with the large 23 rotation, thus inducing a contribution
$\htac$ to $\tac$ which is easily found to be given by
\begin{equation}
  \label{eq:t13cl}
  \sin\htac = \sin\teab
  \frac{\tan\tbc}{\sqrt{\cos^2\theta^e_{12}+\tan^2\tbc}} \simeq
  \sin\teab \sin\tbc \;,
\end{equation}
where $\teab$ is the angle associated to $U^e_{12}$ and the
approximated expression holds for small $\teab$. \Eq{t13cl}
generalizes results in~\cite{Barr:00a,King:03a,Lebed:03a}. By
construction, under the present assumptions the contribution in
\eq{t13cl} is independent of possible additional contributions to
$\tac$. A cancellation among them would be accidental.

The contribution to $\tac$ we obtain this way does not depend on the
many unknowns associated with the model building in the neutrino
sector. In particular, it is independent of the form of the light
neutrino mass matrix; of the mechanism accounting for its smallness;
of the form of the Majorana and Dirac mass matrices in see-saw models.
It is also independent of the origin of the atmospheric mixing angle
(neutrino or charged lepton sector).

On the other hand, $\htac$ does depend on properties of the charged
fermion sector, such as the size of the $\teab$ rotation and the
independence of the entries of $m_E$. In the next section we
illustrate two arguments on the expected size of $\teab$. 

\section{Two arguments on the size of $\teab$}

We first consider a well known, robust ansatz on the structure of the
light $2\times 2$ block of the charged fermion mass matrices. The
ansatz is characterized by a negligible 11 element and by the
approximate equality of the absolute values of the 12 and 21
elements~\cite{Gatto:68a,Ibarra:03a}. Such a
pattern can be accounted for by an elegant non-abelian
symmetry~\cite{Barbieri:97b}.  More important, it leads to the
successful and precise (at the 5\% level) relation
$|V_{us}|=\sqrt{m_d/m_s}$. Furthermore, in a SU(5) grand unified
model, the relations $m^E_{12(21)}=m^D_{21(12)}$, together with the
Georgi-Jarlskog factor 3 in $m^E_{22}=3 m^D_{22}$ (necessary to
account for the muon mass), lead to a second successful prediction:
$m_e/m_\mu=(m_d/m_s)/9$. A prediction for $\teab$ also follows, which
can certainly be considered well motivated, in the light of the above.
The prediction is $\teab = (m_e/m_\mu)^{1/2}\simeq
0.07$~\cite{Barbieri:99a,Ibarra:03a}, leading to
\begin{equation}
  \label{eq:GST}
  \sin\htac \simeq \sqrt{m_e/m_\mu}\sin\tbc \simeq 0.05 \;,
\end{equation}
3--4 times below the present experimental value and therefore within
the reach of future high intensity neutrino beam experiments.

\begin{figure}
\begin{center}
\epsfig{file=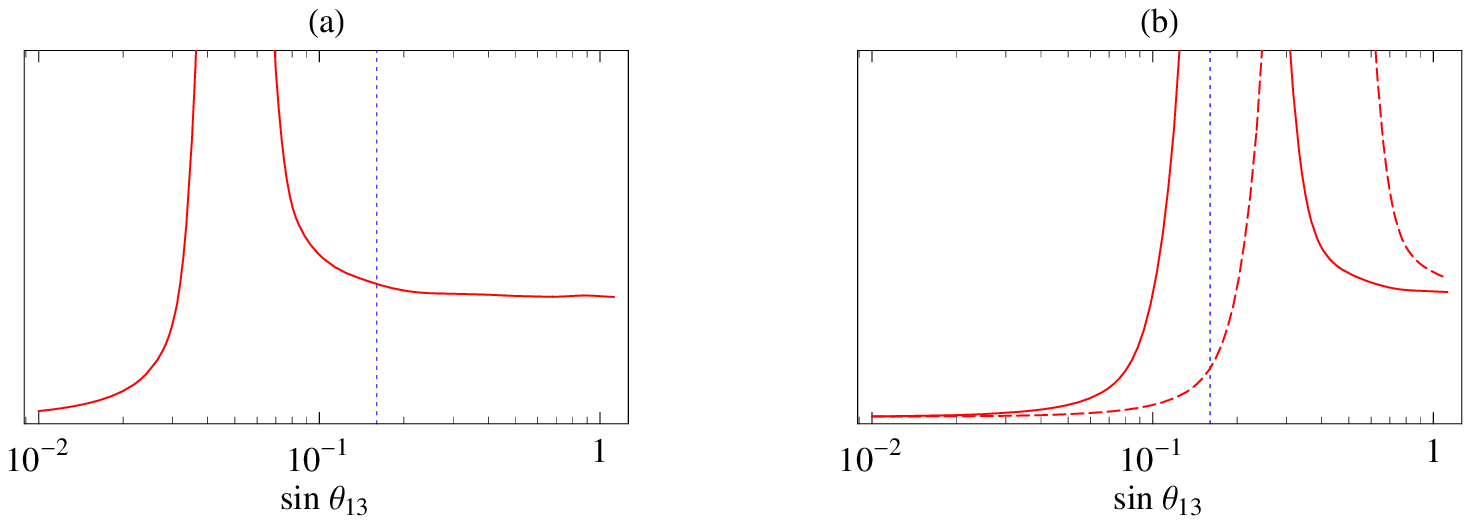,width=\textwidth}
\end{center}
\mycaption{Distribution of $\tac$ obtained by adding a random
   contribution to $\sin\htac = \sqrt{m_e/m_\mu}\sin\tbc$ (a) or
   $\sin\htac = \tan\tbc/2\cos2\tab/\cos\phi$ (b). The vertical dotted
   line represents the present limit. The case $\cos\phi=1$ (solid) and
   $\cos\phi=0.5$ (dashed) are both shown in (b). See comments in the text.}
\label{fig:t13}
\end{figure}

In the presence of additional contributions, \eq{GST} should be
interpreted as a lower bound on $\tac$ that can be evaded if a
cancellation occurs. In order to estimate how small $\tac$ can be made
by such a cancellation, we add to $\sin\htac$ a random contribution
larger than $10^{-4}$, with flat logarithmic distribution and
arbitrary phase. We then obtain a probability distribution for $\tac$,
which is shown in \Fig{t13}a. The peak at $\tac\simeq \htac$ is due to
the small values of the additional contribution (which all give
$\tac\simeq \htac$) and is therefore not particularly meaningful.
More meaningful is the range of $\tac$ in which the probability
distribution is not too much suppressed with respect to the plateau on
the right of the peak. We find that the suppression factor is larger
than 5 for $\tac < 0.02$. In other words, a cancellation leading to
$\tac < 0.02$ is unlikely. \Fig{t13}a is a (mild) generalization of
Fig.~1b in~\cite{Ibarra:03a}, where the additional contribution comes
from a specific structure of the neutrino mass matrix. See
also~\cite{Barr:00a,Lebed:03a}.

\interskip

The case in which the neutrino mass spectrum is of the inverted type
provides a different, purely phenomenological constraint on $\teab$.
Strictly speaking, the argument follows from assuming that the matrix
$\hat U$ in \eq{Lside} is in the form $\hat U = U_{23}(\theta)
U_{12}(\tab=\pi/4)$. This is the typical situation in models leading
to the inverted neutrino mass pattern. In such models, in fact, the
deviation of the neutrino contribution to the solar angle from $\pi/4$
is related to $\dm{21}/\dm{32}$ and turns out to be small compared to
the observed deviation~\cite{King:00a}, barring tunings\footnote{If
  the atmospheric angle comes from the charged lepton sector, the 11
  and 22 elements of the neutrino mass matrix should to be tuned to be
  equal up to a phase. This could be obtained in a non accidental way
  by using a non abelian symmetry.}. Depending on the model, the
atmospheric angle can come from the neutral or the charged lepton
sector, but in both cases $\hat U$ has the above form, up to a
possible further 13 rotation, which we neglect. A sizable $\teab$ is
required in this case to account for the observed value of $\tab$,
significantly different from $\pi/4$~\cite{King:00a,King:02a}. The
relation between the deviation of $\tab$ from $\pi/4$ and the required
$\teab$ involves a physical phase $\phi$~\cite{Barbieri:03a,Me}:
\begin{equation}
  \label{eq:phi}
  \left\{ 
      \begin{aligned}
        \sin\tac &= \sin\theta \sin\teab \\
        \tan\tbc &= \cos\teab \tan\theta \\
        \tan\tab &= \left|\frac{1-\cos\theta\tan\teab
        e^{i\phi}}{1+\cos\theta\tan\teab e^{i\phi}} \right|
      \end{aligned}
      \right. \; .
\end{equation}
Solving for $\teab$, we obtain the following equation for
$\sin\teab/\sqrt{\cos^2\teab+\tan^2\tbc} \equiv x =
\sin\tac/\tan\tbc$:
\begin{equation}
  \label{eq:x}
  \frac{2x}{1+x^2} = \frac{\cos2\tab}{\cos\phi}
\end{equation}
or, neglecting terms quadratic in $x$,
\begin{equation}
  \label{eq:BHR}
  \sin\teab=\frac{\cos2\tab}{2\cos\tbc\cos\phi} \;.
\end{equation}
At the same order, $\tac$ is given by
\begin{equation}
  \label{eq:BHR2}
  \sin\tac = \frac{\tan\tbc}{2} \frac{\cos2\tab}{\cos\phi} \;.
\end{equation}
In explicit models, the phase $\phi$ can be related to the leptonic
CP-violating phase in the standard
parameterization~\cite{Barbieri:03a}.  Numerically, the prediction for
$\tac$ is very close to or beyond the experimental bound, depending on
$\cos\phi$ and on the values used for $\tab$, $\tbc$. In \Fig{t13}b we
show the distribution for $\sin\tac$ we obtain by using the present
distributions for $\tbc$ and $\tab$ and by adding a random
contribution as before. The solid line corresponds to $\cos\phi=1$,
whereas the dashed line corresponds to $\cos\phi=0.5$. In the case
$\cos\phi=1$, the probability is suppressed by a factor of more than 5
(compared with the plateau on the right of the peak) when $\tac < 0.07$.
When $\cos\phi<0.5$, this happens in the whole allowed range.

Note that \eq{BHR} is not compatible with the powerful ansatz
described at the beginning of this section, since it requires a larger
$\teab$ ($\teab\sim 0.25$--$0.30$). In order to avoid a large
contribution to the electron mass, this in turn requires a significant
asymmetry in the 12 block of $m_E$.

\section{$\tab$ from the charged lepton sector (no correlations in $m_E$)}

Let us now consider the possibility that the solar angle originates
predominantly in the charged lepton sector. We first consider again
the case in which the entries in $m_E$ are independent. Then $\tab$
must originate from the $U^e_{12}$ factor in~(\ref{eq:Lside}). Since
the required size of the $\teab$ rotation is now quite large,
\begin{equation}
  \label{eq:te12}
  \sin\teab = \frac{\sin\tab}{\cos\tbc} \simeq 0.8 \;,
\end{equation}
we can expect the induced contribution to $\tac$ to be large. Indeed,
in the limit in which the matrix $\hat U$ in~(\ref{eq:Lside}) consists
of a pure 23 rotation, we have
\begin{equation}
  \label{eq:t13cl2}
  \sin\htac = \tan\tbc\tan\tab \simeq \text{0.6--0.7},
\end{equation}
4--5 times above the experimental limit. Therefore, the matrix $\hat
U$ should contain an additional 13 or 12 rotation factor cancelling
most of $\htac$. We will refer to this problem as the ``$\tac$ tuning
problem''. In this scenario, one of the few hints on the origin of
lepton mixing (the smallness of $\tac$) would be an accident. On the
other hand, the degree of cancellation needed to bring $\tac$ below
the experimental limit is mild. A potentially more serious
tuning problem comes from the $m_e\ll m_\mu$ hierarchy, at least in
SU(5) models, as we now see.

Let us call $\teabc$ the right-handed rotation involved in the
diagonalization of the light $2\times 2$ block of the charged lepton
matrix (after diagonalization of the 23 block). The $e/\mu$ mass ratio
is then given by
\begin{equation}
  \label{eq:emu}
  \frac{m_e}{m_\mu} \simeq \left| \frac{m^E_{11}}{\hat m^E_{22}} -
  \tan\teab \tan\teabc \right| \;,
\end{equation}
where $\hat m^E_{22}$ is the 22 entry of $m_E$ after diagonalization
of the 23 block. As a consequence, we expect $m_e/m_\mu$ to be of the
order of $\tan\teab \tan\teabc\sim \tan\teabc$ or larger, barring
cancellations. On the other hand, $\teabc$ is related to the
left-handed 12 rotation in the down quark sector $\theta^d_{12}$
(analogous of $\teab$) by GUT relations: $\theta^d_{12}=C \teabc$,
where $C$ comes from possible SU(5) Clebsch-Gordan factors. We also
know that $\theta^d_{12}$ contributes to the CKM element $V_{us}$.
Indeed, if the up quark mass is to be naturally small, the up quark
contribution to $V_{us}$ must be subdominant. That is because the up
quark matrix is symmetrical in SU(5) and $\theta^u_{12}\sim |V_{us}|$
would give a large contribution to $m_u/m_c$ through a relation
similar to~(\ref{eq:emu}). We then have $\theta^d_{12}\simeq |V_{us}|$
and a contribution to $m_e/m_\mu$ of order $|V_{us}|/C$. In minimal
SU(5), this would be about 50 times larger than the measured value.
Even in the presence of a (plausible) Clebsch-Gordan factor, the
necessary tuning is still larger than that involved in the $\tac$
tuning problem.  We refer to this problem as the ``$m_e$ tuning
problem''.

In summary, in absence of correlations in $m_E$, generating the solar
mixing angle from the charged lepton sector requires cancellations in
the determination of $\theta_{13}$ and, in SU(5) models, in the
determination of the electron (or up quark) mass.

\section{$\tab$ from correlations in $m_E$}

Let us now consider the case in which the entries of the charged
lepton mass matrix are correlated. It has recently been
shown~\cite{AFM} that such correlations can be naturally induced in a
Froggat--Nielsen context by the dominance of a heavy vector-like lepton
exchange (analogous to the single right-handed neutrino dominance
scenario in the neutrino sector~\cite{Smirnov:93b}).  More precisely,
the approximate vanishing of the determinant of the $x_{ij}$
coefficients in
\begin{equation}
  \label{eq:AFM}
  m_E \propto 
  \begin{pmatrix}
    a\epsilon' & b\epsilon' & \ord{\epsilon'} \\
    x_{21} \epsilon & x_{22} \epsilon & \ord{\epsilon} \\
    x_{31} & x_{32} & 1
  \end{pmatrix}
\end{equation}
($\epsilon'\ll\epsilon\ll 1\sim a,b,x_{ij}$) has been identified as a
condition for the large $\tab$ (and $\tbc$) to originate in a natural
way from $m_E$. The correlations in $m_E$ translate in fact in a
relation between the 12, 13, 23 rotations in \eq{order} cancelling the
physical $\tac$ angle. This can easily be seen by observing that,
unlike the case in which correlations are absent, the proper way to
diagonalize $m_E$ is by performing the left-handed 12 rotation first,
followed by the 23 rotation. 

The form of $m_E$ in~(\ref{eq:AFM}) is not unique. For example, the
pattern
\begin{equation}
  \label{eq:mo}
  \begin{pmatrix}
    \epsilon' & 0 & x_{13} \epsilon \\
    0 & \epsilon' & x_{23} \epsilon \\
    x_{31} & x_{32} & 1
  \end{pmatrix} \;,
\end{equation}
with $|x_{13} x_{32} - x_{31} x_{23}|\ll 1$, also gives a large solar
angle in a natural way. However, it is not compatible with SU(5) (see
below) and it is harder to obtain from the model building point of
view (see however~\cite{Kuchimanchi:02a}). Another possibility is that
correlations are not present in the initial form of $m_E$ but they are
induced by the first steps of the diagonalization in~(\ref{eq:order}).
In the following we will concentrate on the possibility in \eq{AFM}.

While the $|x_{21} x_{32} - x_{31} x_{22}|\ll 1$ condition on the
coefficients in~\eq{AFM} does solve the $\tac$ problem discussed in
the previous section, the additional condition
\begin{equation}
  \label{eq:condition}
  |a x_{22} - b x_{21}|\ll 1
\end{equation}
must be imposed in order to ensure that the electron mass and the up
quark mass are naturally small in SU(5) grand unified models (namely
to ensure that the $m_e$ problem illustrated above is also solved).
The only possibility to escape the condition~(\ref{eq:condition}) is
boosting the 13 entry in~(\ref{eq:AFM}), namely imposing the
alternative condition $|m^E_{11}|$, $|m^E_{12}|\ll |m^E_{13}|$.

The argument goes as follows. In SU(5) the light block of the up quark
mass matrix is symmetric. As a consequence, the up quark contribution
to $|V_{us}|$ must be subdominant, as discussed in the previous
Section, and we can identify $|V_{us}|$ with its down quark
contribution. We then obtain, e.g.\ in minimal SU(5),
\begin{equation}
  \label{eq:error}
  \frac{m_e}{m_\mu}\cos\theta_{23} \sim \left| V_{us} \frac{a x_{22} - b
  x_{21}}{ax_{21}+bx_{22}} \right| \;.
\end{equation}
The condition~(\ref{eq:condition}) follows from $|V_{us}|\gg
m_e/m_\mu\cos\theta_{23}$. We have used $\teabc\simeq |V_{us}|$ and we
have neglected the contribution of the 13 entry in~(\ref{eq:AFM}) to
$\teabc$ that arises due to the 23 rotation. If that entry is larger
than $\ord{\epsilon'}$, its contribution to $\teabc$ and $V_{us}$
dominates and the connection with $m_e/m_\mu$ is lost, thus leading to
the alternative possibility $|m^E_{11}|$, $|m^E_{12}|\ll |m^E_{13}|$.

The above can be rephrased by using the example in Section 1
of~\cite{AFM}. Assume that the lepton mixing matrix is in the form $U
= U_{23}(\tbc = \pi/4)U_{12}(\theta)$. Then, using the notation of
eq.~4 in~\cite{AFM},
\begin{equation}
  \label{eq:AFM2}
  m_E = V_e 
  \begin{pmatrix}
    c\, m_e & s\, m_e & 0 \\
    s/\sqrt{2}\, m_\mu & -c/\sqrt{2}\, m_\mu & m_\mu/\sqrt{2} \\
    -s/\sqrt{2}\, m_\tau & c/\sqrt{2}\, m_\tau & m_\tau/\sqrt{2}
  \end{pmatrix} \;.
\end{equation}
Given the relation between $(V_e)_{12}$ and $V_{us}$, taking $V_e\sim
1$ is not a good approximation in SU(5). For instance, in minimal SU(5)
we have
\begin{equation}
  \label{eq:AFM3}
  m_E \simeq
  \begin{pmatrix}
    V_{us} s/\sqrt{2} m_\mu & -V_{us} c/\sqrt{2} m_\mu & V_{us} m_\mu/\sqrt{2} \\
    s/\sqrt{2} m_\mu & -c/\sqrt{2} m_\mu & m_\mu/\sqrt{2} \\
    -s/\sqrt{2} m_\tau & c/\sqrt{2} m_\tau & m_\tau/\sqrt{2}
  \end{pmatrix} \;.
\end{equation}
Additional contributions to the first and second rows proportional to
$V_{ub}$ and $V_{cb}$ might also be relevant, but they have been
omitted in~(\ref{eq:AFM3}). 

From the model building point of view, the additional
condition~(\ref{eq:condition}) does not represent an additional
challenge. The same mechanism accounting for the correlation in the
lower left block of $m_E$ may well account also for the correlation in
the upper left block. This is illustrated by the explicit SU(5) model
in Appendix A of~\cite{AFM}. In the light of \eq{AFM3}, we remark
however that one might expect the first row of $m_E$ to be suppressed
by only a factor $|V_{us}|$ compared to the second row.

\section{Summary}

We have shown that a charged lepton contribution to the solar neutrino
mixing induces a contribution to $\tac$, barring
cancellations/correlations, which is independent of the model building
options in the neutrino sector. We have illustrated two robust
arguments for that contribution to be within the expected sensitivity
of high intensity conventional neutrino beam experiments. The
corresponding expectations for $\tac$ are shown in \Fig{t13}. In
particular, in the case in which the neutrino sector gives rise to a
maximal solar angle (the natural situation if the neutrino spectrum is
inverted, barring non accidental tunings) we have given a simple
analytical expression for the induced contribution to $\tac$ taking
into account the dependence on a physical phase. The numerical value
of $\tac$ turns out to be very close or exceeding the experimental
bound, depending on the precise value of the solar and atmospheric
mixing angles, the value of the phase, and the presence of additional
contributions. We have also discussed the possibility that the solar
angle originates predominantly in the charged lepton sector. In the
case in which no correlations are forced in the charged lepton mass
matrix, we have shown that this possibility faces two fine-tuning
problems, one in the determination of $\tac$ and one in the
determination of the electron (or up quark) mass in SU(5) models.  We
have also considered the case in which correlations are present in the
charged lepton mass matrix and, in the context of the possibility
discussed in~\cite{AFM}, we have identified the additional conditions
that allow to solve the $m_e$ tuning problem in SU(5) models.


\end{document}